\documentclass[fleqn,10pt]{wlscirep}
\usepackage[utf8]{inputenc}
\usepackage[T1]{fontenc}
\usepackage{lineno}
\title{Ageing related states of complex network formation in areca nuts}

\author[1,2,3,$\ast$]{KV Chinmaya}
\author[1,2,3,$\S$]{Moumita Ghosh}
\author[1,2,3,$\ddag$,$\ddag \ddag$,$\dag$,$\ast$]{Siddharth Ghosh}
\affil[1]{Department of Science, Open Academic Research Council, Kolkata, India}
\affil[2]{Department of Science, Open Academic Research UK CIC, Cambridge, UK}
\affil[3]{International Center for Nanodevices, Centre for Nano Science and Engineering, Indian Institute of Science, Bengaluru, India}
\affil[$\ddag$]{Department of Applied Mathematics \& Theoretical Physics, University of Cambridge, Cambridge, UK.}
\affil[$\ddag \ddag$]{Maxwell Centre, University of Cambridge, Cambridge, UK.}
\affil[$\dag$]{St John's College, University of Cambridge, Cambridge, UK.}
\affil[$\ast$]{corresponding authors: chinmay@openacademicresearch.org and siddharth@openacademicresearch.org}
\affil[$\S$]{Current affiliation: Advanced Technology Group of FEI Electron Optics, Thermo Fischer Scientific, Eindhoven, The Netherlands}

\begin{abstract}
 Complex pattern formation is an essential characteristic of plants and their ageing, growth, and evolution. 
   Perception of these patterns is an intrinsic nature of plant-dependent animals for coexistence.
    Areca nut consisting of complex patterns is considered to be addictive for humans and has increased adverse health effects. 
    However, no critical study is performed on the complex pattern of the areca nut.
    A large number of areca nuts has been studied since 2017 to develop a low-cost tool for the LMICs to categorise areca nuts.
    We present the first finding to identify similarities among complex networks of differently aged areca nuts by investigating the internal patterns of randomly chosen nuts from the same age group. 
   We developed a smartphone camera-based high-resolution measurement with comprehensive biophysical mathematics and a quantum mechanical concept called density of states (DOS). 
    We found that the DOS can provide a unique coefficient to represent age and ageing together using a single number. 
    The average of these single numbers for less aged nuts and highly aged nuts are 4.9 and 3.8, respectively. 
   If fruit looks aged from its external morphology as well as internal morphology, our method identifies the intrinsic similarities among the ageing networks without implementing any computationally expensive search algorithm. 
   We show clear evidence of the diversity of ageing from relative and absolute colour vision perspectives.
    We have also conducted further analyses of local DOS, Fourier decomposition, correlation study, spectral decomposition, and structural similarity index.  
\end{abstract}

\begin{document}

    \flushbottom
    \maketitle
     \thispagestyle{empty}
    % \begin{center}
    %       \includegraphics[width=0.8\textwidth]{TOC.png}
    % \end{center}
    % \newpage
    %  \textbf{
     
        The evolution of plants in the kingdom eukaryotic Plantae phenotypically expresses complex patterns\cite{boller2009innate,yuan2021pattern}.
        These patterns are fascinating to the animal kingdom for the coexistence of species by mimicry, adaptive camouflage, and taxis \cite{wehner1967pattern,merilaita2017camouflage,pannell2016mimicry,jin2019prepatterning,eacock2019adaptive}.
        Plant-derived patterns have created new branches of mathematical physics\cite{langlois1990attractive,minarova2014fibonacci} since many of these complex patterns can not be explained by conventional mathematical functions \cite{jardine1971new,weinberg1972general,gfeller2007spectral,bamberger2015dealing,menichetti2015multiscale}.
        Life on earth shows significant evidence of chaotic patterns\cite{hoffman1966nature,1995fractal,nathan2000spatial,2007Barbosa1139,niessing2010olfactory,2010Katifori,kumar2014pulsatory,stoddard2014pattern,jhawar2020noise,elmarakeby2021biologically}.
        Necessity of these patterns motivates us to study statistical mechanics of complex systems, which has drastically progressed to address these problems\cite{albert2002statistical,skardal2014optimal,charbonneau2012universal,burda2020ageing}, namely, unifying approach of Turing patterns \cite{kirkwood1977evolution,kondo2010reaction,zheng2016identifying,hearn2019turing}.
        One of the unresolved questions is if these complexities arise from the fundamental self-assembly of matters and the role of defects in them\cite{kirkwood1977evolution,hynninen2007self,hennig2011nature,ghosh2016atomic,zhou2018real}.
        %However, an obvious question remains, 
        We have identified one such complex pattern formation of ageing in areca nuts (\textit{Areca catechus}). 
        These patterns arise due to complex interactions of alkaloids and other biochemicals \cite{raghavan1958arecanut, huang2012detection, danti2012segmentation}.
        We developed a smartphone camera-based high-resolution technique with a set of comprehensive biophysical mathematics complementing density of states (DOS). 
        The internal morphology of areca nuts possesses two distinctive regions in the kernel of white and brown.
        Within same age groups, ageing is more in some nuts than the others. 
        We found that the DOS can provide a unique coefficient to represent age and ageing together using a single number.
        Here, we have performed a comprehensive analysis of the morphological network formation in areca nuts of the same off-springs with the same age groups from relative and absolute colour vision perspectives.
        Our method identifies the intrinsic similarities among the ageing networks without implementing any computationally expensive search algorithm. 
        These quantitative analysis of diversity of ageing \cite{kirkwood1977evolution,jones2014diversity,freund2019untangling,cohen2020aging} bring insight in animal ageing, evolutionary psychology of complex pattern formation for phenotypic expression of age\cite{perez2017maternal}, the chemistry of health \cite{whoareca} for areca nuts consumption, computer vision for developing efficient algorithms, and commonality among fingerprint patterns\cite{KUCKEN200571}, nanomaterials self-assembly, and quantum fluids.\cite{1998fabry,feng2019correlations,edmonds2020vortex}
        % }
    
%\noindent\textbf{MAIN}

% Complex systems face stagger setback in detecting randomness and pattern.  
% The answers to them give us an insights of how nature reckons these complex patterns.
% The system might looks like a complicated problem but is it actually a complex?
% On what basics we say it is complex?, is it our mind saying complex or the actual problem is complex like biological evolution?

\vspace{2mm}
Complex evolution of ageing in biological system faces stagger setback in detecting randomness and pattern due to the interaction of many units. 
Decomposing the complex network of a system with a high-throughput method leads, fast understanding of complex system properties.
We present the first concept of fundamental understanding of ageing by decomposing the complex networks in an overlooked biological system of areca nuts, which have been consumed by a large number of human population across the globe\cite{whoareca, paulino2017areca}.
Here, the state of the art describes the unification of ageing related complex networks in areca nut.

\subsection*{Experimental overview of ageing-related growth}

\begin{figure*}[htp]
        \centering
        \includegraphics[width=0.6\textwidth]{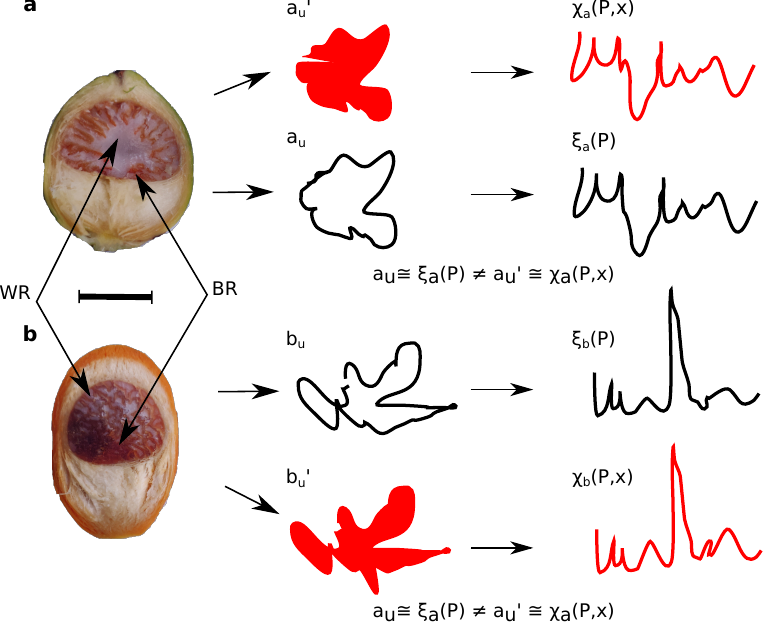}
        \caption{\textbf{Schematic flowchart of a novel  principle of ageing related growth patterns of areca nut}. \textbf{a}, Cross-section of a unripened areca nut with green outer shell. 
        \textbf{b}, Cross-section of a ripened areca nut with orange outer shell. 
        It reveals the internal patterns of the nut constituted with white (WR) \ref{figure:1}\textbf{a} and brown (BR) Figure \ref{figure:1}\textbf{b} regions, shown in both the nuts. 
        Scale bars in both cases are 11.5 mm.  \textbf{a$_u^\prime$, a$_u$} and \textbf{b$_u^\prime$, b$_u$} are the schematic of the mathematical depiction of both unripened and ripened nuts respectively.}
       \label{figure:1}
    \end{figure*}
    
We show two exemplary areca nuts' cross-sections in Figure \ref{figure:1}. 
They were captured using a standard smartphone; technical details are available in the method section.
The outer shell of the unripened nut is green in Figure \ref{figure:1}\textbf{a} and brownish-orange colour in case of the ripened one in Figure \ref{figure:1}\textbf{b}. 
Both are approximately 90 days old and harvested on the same day from the same branch of same tree in the month of March. 
The major axis of the areca nut testa vary from 1 cm to 3.5 cm and the minor axis from 1 cm to 2.5 cm.
%, and the ratio between them ranges from 1:1 to 1:1.5. 
The testa of the areca nut contains brown and white regions with the pattern of complex networks as shown in Figure \ref{figure:1}\textbf{a} and Figure \ref{figure:1}\textbf{b}. 
Since the age-dependent evolution of the brown region begins from the edges of the testa with strong peripheral localisation, we can consider that the transportation of nutrients into the nut is following the fibre networks in the husk \cite{kramer1995water,kurant2006layered,gomez2007paths,son2009dynamics,charbonneau2012universal,skardal2014optimal,muralidhar2019study} or a reaction diffusion mechanism \cite{kondo2010reaction}.
In Figure \ref{figure:1}\textbf{a$_u^\prime$}, Figure \ref{figure:1}\textbf{a$_u$}, Figure \ref{figure:1}\textbf{b$_u$}, and Figure \ref{figure:1}\textbf{b$_u^\prime$}, we show the principle of ageing related growth patterns of areca nuts.  
We investigated the time-dependent density of states associated with these two types of phenotype expressions to identify the morphological change during the ageing. 
%Structure of the paper by explaining figure 1.

\subsection*{States of complex network of ageing}

We performed smartphone based imaging to identify the states of complex network of ageing in areca nuts. 
The nuts have a high signal to noise ratio colour contrast to distinguish between the white and brown regions as shown in Figure \ref{figure:1}\textbf{a} and Figure \ref{figure:1}\textbf{b}.
We have given unique identification symbol for each areca nuts to grasp the analysis in later sections.

\begin{figure*}[htp]
    \includegraphics[width=1\textwidth]{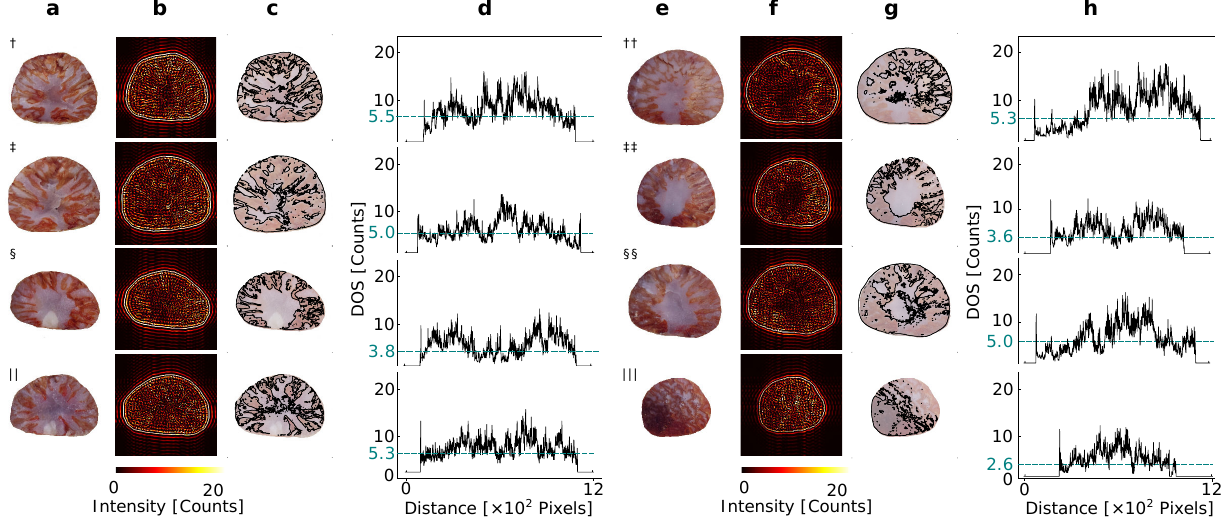}
    \caption{\textbf{Smartphone camera images of areca nuts, their edge-detection and density of states (DOS).} \textbf{a}, 
    Unripened nuts -- original unripened areca nuts respectively. 
    \textbf{b}, HPF from 2D-FFT of \textbf{a}. \textbf{c}, overlay of edge detection from unfiltered data of \textbf{a} using black data points. \textbf{d}, DOS of canny edge detection vs distance. 
    Ripened nuts -- \textbf{e}, original ripened  nuts followed by its respective analysis. \textbf{f}, HPF from 2D-FFT of \textbf{e}. \textbf{g}, the overlay of canny edge detection from \textbf{e}. \textbf{h}, DOS of canny edge detection vs distance.}
    \label{figure:2}
\end{figure*}

In Figure \ref{figure:2}\textbf{a} we show the original unripened nuts. 
We analysed the spatial frequency of two regions of both age groups, initially, using 2D FFT analysis (see supplementary Figure 1) to investigate the complex network and filter out the high-frequency domains as shown in Figure \ref{figure:2}\textbf{b}. 
We found strong ringing artefacts as shown in supplementary Figure 2. 
We could not effectively extract the network of brown and white regions due to these artefacts \cite{reeves2005fast}. 
We observed some fine features, which were later confirmed (from other correlative studies) to be responsible for the alkaloids.   
To identify the network, we detected the edges between the brown and white regions using the Canny edge detector (see Methods for the parameters). 
We distinguished two morphological regions for unripened in Figure \ref{figure:2}\textbf{c}. 
The morphological network in Figure \ref{figure:2}\textbf{c} were obtained effectively due to the prominent colour based edge detection between white and brown regions. 
Although visually these differences were prominent in some cases, the algorithm could not detect several edges, letting a few networks fall into discontinuities.

To approach a single number of quantification and quantitative comparison of the distribution of edges, we introduced the concept of DOS.
The DOS of a nut describes the density of detected edges' states that are occupied within an image in terms of pixels. 
The total number of density of states in each nut is calculated following the equation \ref{eq:3} in the Method section.
Figure \ref{figure:2}\textbf{d} shows the dependency between the DOS and distance of four unripened nuts.  
We observe some strong peaks (`with bunching') and valleys, which are distinguishable from the mean values. 
The morphological edges seem to follow a random trend in contrast to a periodic structures (for example, supplementary Figure 3). 
We further quantified the mean average precision (mAP) of DOS that is the average of DOS with respect to the total number of pixels as shown with teal coloured numbers in Figure \ref{figure:2}\textbf{d}.
The mAP ranges from $3.8$ to $5.5$ counts for unripened nuts calculated using the Equation \ref{eq:9}.
We have used the same technique for ripened nuts in Figure \ref{figure:2}\textbf{e} followed by the 2D-FFT in Figure \ref{figure:2}\textbf{f}, Canny edge detection in Figure \ref{figure:2}\textbf{g}, and the dependency between the DOS and distance of four ripened nuts in Figure \ref{figure:2}\textbf{h}. 
The mAP ranges from $2.6$ to $5.3$ counts for ripened nuts which is collectively same as unripened nuts. 
The average of these single numbers for less aged nuts and highly aged nuts are 4.9 and 3.8, respectively.
The variation of the mAP is not high in both type of the nut. 
In future, one can consider this as an order parameter of complexity in the DOS.

\subsection*{Decomposition of complex networks of ageing}

\begin{figure*}[htp]
    \centering
    \includegraphics[width = 1\textwidth]{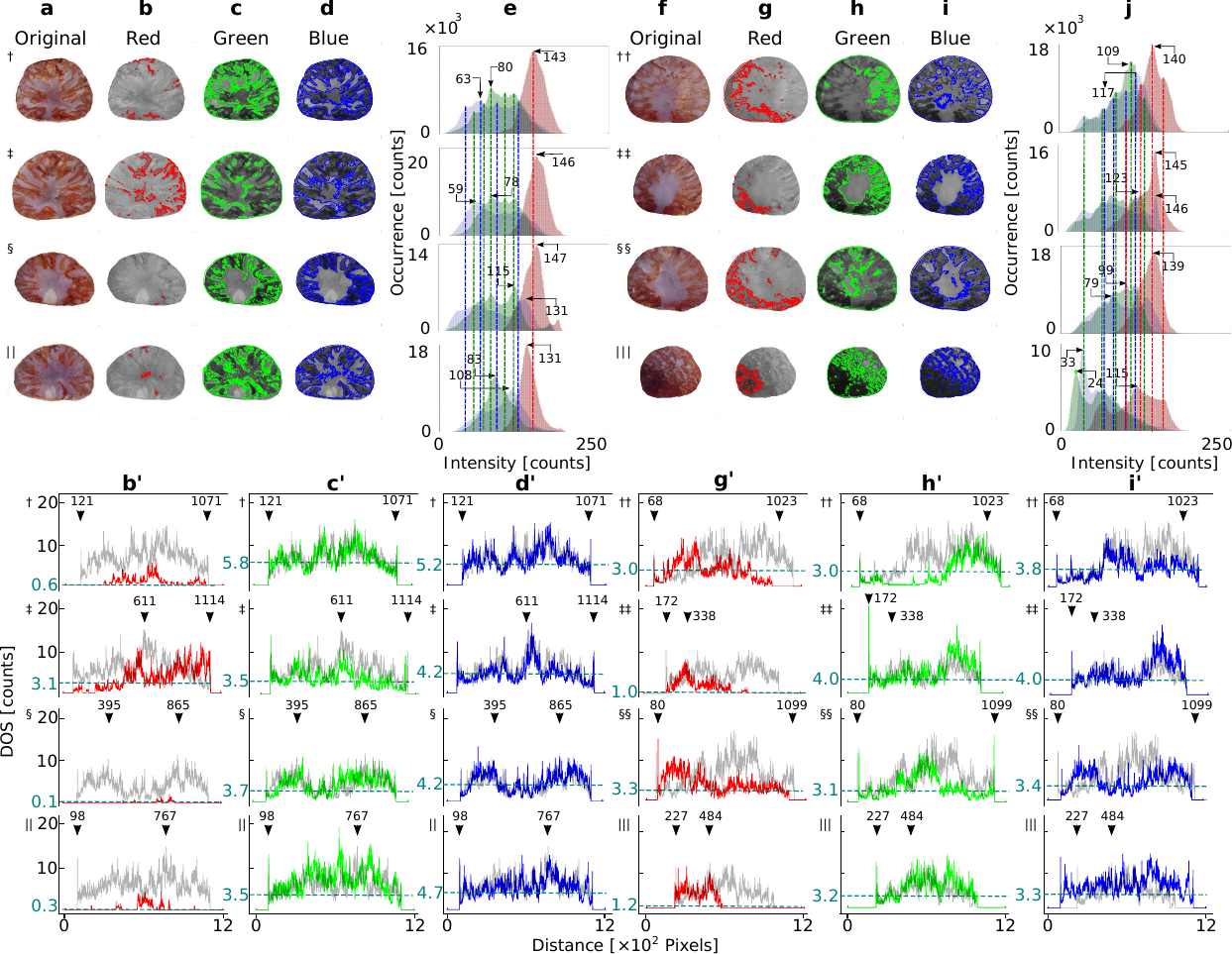}
    \caption{\textbf{Colour decomposition of unripened and ripened areca nuts.} 
    \textbf{a}, Unripened nuts followed by its analysis. 
    \textbf{b}, Red decomposition image which has very minimum edges. 
    Unripened nuts -- \textbf{b$^\prime$}, LDOSs of areca nuts from `b' -- variation against distance.
    The mean average precision (mAP) is introduced with the dotted lines with the number. 
    \textbf{c}, Green decomposition and its corresponding states. 
    \textbf{c$^\prime$},  LDOSs of areca nut from `c' -- variation against distance. 
    The mAP of green LDOSs is much higher than red LDOSs. 
    \textbf{d}, Blue decomposition and its corresponding LDOSs in b$^\prime$.
    \textbf{e}, RGB decomposition histograms of the unripened nuts where red, green, and blue spectral shifts can be identified with respect to the first nut as reference using the guiding dashed lines. 
    \textbf{f}, Ripened nuts followed by its analysis. 
    ripened nuts -- \textbf{g}, Red decomposition image which has very minimum edges. 
    \textbf{g$^\prime$}, LDOSs of areca nut from `g' -- variation against distance. 
    mAP of red LDOSs is very low. 
    This is evident for a few states in it. 
    \textbf{h}, Green decomposition and its corresponding states. 
    \textbf{h$^\prime$}, LDOSs of areca nut from `h' -- variation against distance. 
    The mAP of green LDOSs is much higher than red LDOSs. 
    \textbf{i}, Blue decomposition and its corresponding LDOSs in i$^\prime$. 
    \textbf{j}, RGB decomposition histograms of the ripened nuts where the spectral shifts can be identified using the guiding dashed lines, like \ref{figure:3}\textbf{e}.
    The arrows in each sub figure from b$^\prime$ to i$^\prime$ indicates the position of peak from a distance (pixels)}
    \label{figure:3}
\end{figure*}

We further investigated fine features of the network and decompose the original images in Figure \ref{figure:3}\textbf{a} into different colour channels to quantitatively deduce whether the random trend is also followed in the decomposed networks. 
Figure \ref{figure:3} shows the red ($\lambda = 690$ nm in \ref{figure:3}\textbf{b}), green ($\lambda = 510$ nm in \ref{figure:3}\textbf{c}), and blue ($\lambda = 440$ nm in \ref{figure:3}\textbf{d}) colour decomposition. 

We identified wavelengths of the primary colours from Newton's colour triangle to wavelength relationship (see supplementary information).
We obtain the spectral distribution by quantifying the occurrence of the aforementioned three basic colours with respect to their intensities from the original colour images in contrast to three classes of LDOSs.
In Figure \ref{figure:3}\textbf{e}, we show RGB decomposition histograms of original colour image of unripened areca nuts as shown in Figure \ref{figure:3}\textbf{a}. 
We have marked the first nuts' RGB intensity peak as a standard peak, 143 counts for red, 80 counts for green, and 63 counts for blue, then we extrapolated to the rest of all three RGB peaks (shown in red, green and blue dotted lines).  
We saw that the first nut has a single red peak, while the third nut has three more.
Green and blue have five and four standard peaks, respectively, which are less frequent than red.
Similarly for ripened nuts, the original ripened nuts in Figure \ref{figure:3}\textbf{f}, splitted in three colour channels shown in Figure \ref{figure:3}\textbf{g} - \textbf{i}. 
In Figure \ref{figure:3}\textbf{j}, we show RGB decomposition histograms of original colour image of ripened areca nuts as shown in Figure \ref{figure:3}\textbf{f}.
Four peaks in red, five summits in green, and four peaks in blue were found here.
Like unripened nuts, all green and blue peaks have a lower frequency than red.
We also noticed that some of the red, green, and blue peaks cross other channels' peaks.

We perform canny edge detection of each colour channels, which corresponds to the local density of states (LDOSs) of the total DOS --- the detected edges surround the regions where that colour is unavailable.  
The red LDOSs are localised and restricted to a particular regions only as seen in the Figure \ref{figure:3}\textbf{b} for unripened nuts.
The red channels possess low contrasts, and the network can not be seen distinctively.
The red LDOSs are not evenly dispersed throughout all of the nuts. 
They are  primarily localised at the edges with a few cases at the centre.   
The networks can not be seen distinctively in the red channels and the red LDOSs are low in counts as shown in Figure \ref{figure:3}\textbf{b$^\prime$}, where the mAP ranges from $0.1$ to $3.1$ counts for unripened nuts.  
We have overlaid the DOS plot of each nut to its corresponding LDOSs in the background. 
In the Figure \ref{figure:3}\textbf{b$^\prime$} DOS has the predominance signal with respect to red LDOSs. 
The green and blue channels posses high concentrations of states than red unripened nuts shown in Figure \ref{figure:3}\textbf{c} and Figure \ref{figure:3}\textbf{d} respectively. 
The distribution of corresponding states are shown in Figure \ref{figure:3}\textbf{c$^\prime$} and Figure \ref{figure:3}\textbf{d$^\prime$}, the mAP ranges from $3.5$ to $5.8$ counts for green and $4.2$ to $5.2$ counts for blue LDOSs respectively.   
The ripened nuts also show similar behavioural states of red, green, and blue as shown in Figure \ref{figure:3}\textbf{g} and Figure \ref{figure:3}\textbf{h}, and \ref{figure:3}\textbf{i} respectively.
The distribution of corresponding states are shown in Figure \ref{figure:3}\textbf{g$^\prime$} and Figure \ref{figure:3}\textbf{h$^\prime$},
and Figure \ref{figure:3}\textbf{i$^\prime$}, the mAP ranges from $3.5$ to $5.8$ counts for red and $1.0$ to $3.3$ counts for red, $3.0$ to $4.0$ counts for green, and $3.3$ to $4.0$ for blue respectively. 

We find in Figure \ref{figure:3}\textbf{g} and Figure \ref{figure:3}\textbf{h} of the first sample, the LDOSs of red and blue channels seems to be anti-correlated or there is no correlation, Figure \ref{figure:3}\textbf{g$^\prime$} and Figure \ref{figure:3}\textbf{h$^\prime$} of the same plot gives a broader idea of the distribution of LDOSs. 
Figure \ref{figure:3}\textbf{g$^\prime$} has a surge at the beginning of the left-hand side of the plot.
We obverse that number of the edge detected LDOSs are mostly concentrated in the left side of Figure \ref{figure:3}\textbf{g}. 
A significantly opposite trend is present in the green channel Figure \ref{figure:3}\textbf{h} and Figure \ref{figure:3}\textbf{h$^\prime$} of the first image sample. 
The surge in the plot is on the right-hand side. 
We observe the mAP of both LDOSs is 3. 
The mAP remains the same but the variation in LDOSs is always there. 
This motivated us to calculate the 2D-normalised cross-correlation coefficient for LDOSs. 
The results showed there is no strong correlation between any of the LDOSs (see supplementary Figure 4 and Figure 5).
The results of 2D-normalised cross-correlation show the convincingly high correlation between green and blue LDOSs. 
The 2D-normalised cross-correlation coefficient is larger when the two image matrices of green and blue are correlated than the red.
In all the results, the local red LDOSs are not present in all eight nuts, they are present globally. 
Our assumption is evident when we see flat distribution with of red channel histogram in both types of age groups. 
The green and blue contributions in the histogram are low. 
More peaks in green and blue are contributing LDOSs.
From the visualisation perspective, the red LDOSs are less because of few peaks in the red histogram. 
In contrast to the unripened nuts, a single peak of red histogram leads to low mAP values in the LDOSs but four peaks of red in ripened nuts leads to better mAP compared to unripened LDOSs. 

\subsection*{Colour-based local density and fine features of the complex networks}

\begin{figure*}[t!]
    \centering
    \includegraphics[width=1\textwidth]{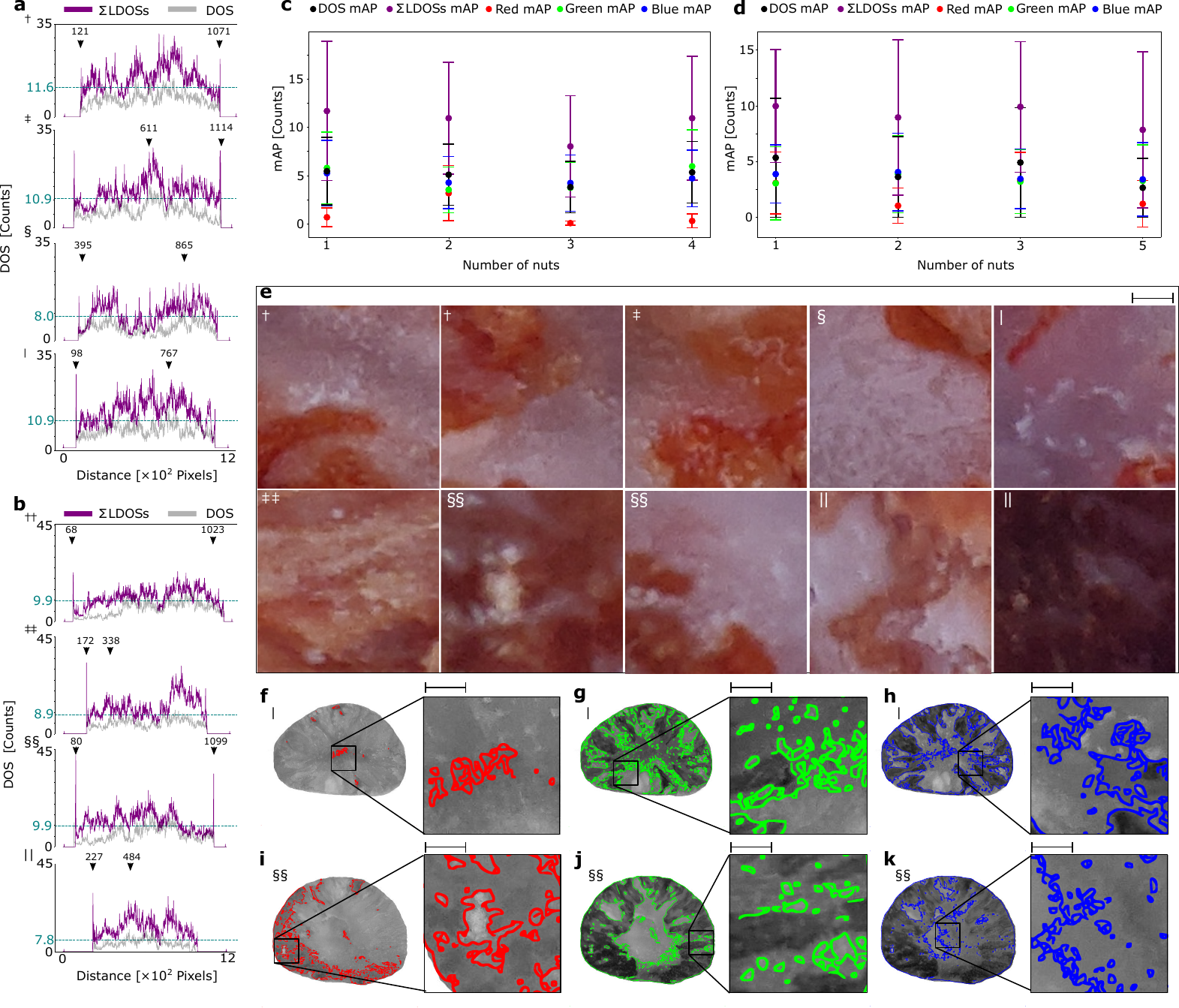}
    \caption{\textbf{Local density of states in contrast to density of states and fine features of the networks.} 
    \textbf{a}, Sum of LDOSs for unripened areca nuts overlaid on DOS.
    \textbf{b}, Sum of LDOSs of ripened areca nuts overlaid on DOS
    \textbf{c}, Unripened nuts -- variation of mean LDOS mAPs and DOS maps and their standard deviation.
    \textbf{d}, Ripened nuts -- variation of mean LDOS mAPs and DOS maps and their standard deviation.
    \textbf{e}, Fine features or detailed morphology of the complex networks in the raw images.
    Scale bars in \textbf{e} is 9.1 mm.
    Colour channels of LDOS and its fine features at different locations of  --
    \textbf{f}, Red channel an unripened nut.
    \textbf{g}, Green channel of an unripened nut.
    \textbf{h}, Blue channel of unripened nut.
    \textbf{i}, Red channel of a ripened nut.
    \textbf{j}, Green channel of a ripened nut.
    \textbf{k}, Blue channel of a ripened nut.
    Scale bars in \textbf{f} - \textbf{k} are 9.1 mm.}
    \label{figure:4}
\end{figure*}

In Figure \ref{figure:4}\textbf{a}, we show the sum of all LDOSs of unripened nuts taken from Figure \ref{figure:3}\textbf{b$^\prime$}, Figure \ref{figure:3}\textbf{c$^\prime$}, and Figure \ref{figure:3}\textbf{d$^\prime$}. 
Violet colour represents the sum of LDOSs and transparent black represents the DOS. 
Similarly, for ripened, Figure \ref{figure:4}\textbf{b}, we show the sum of all LDOSs of ripened nuts taken from Figure \ref{figure:3}\textbf{g$^\prime$}, \ref{figure:3}\textbf{h$^\prime$}, and \ref{figure:3}\textbf{i$^\prime$}.

The total sum of LDOSs of the nuts is not equal to the total DOS. 
This is an important finding from decomposing the original images, we found more states in LDOSs than the DOS.  
In order to find out the variation of mAP in the LDOSs, we plotted the variation of mAP with respect to nuts.
Figure \ref{figure:4}\textbf{c} and Figure \ref{figure:4}\textbf{d} are the mAP of LDOSs of unripened and ripened nuts. 
Standard deviation is calculated using the Equation \ref{eq:12} and plotted on the mAP. 
mAP is obtained from both unripened and ripened nuts. 
A chaotic behaviour is observed in this plot.
The contributions of red mAP from the LDOSs are less.
Decomposing DOS also helped us to visualise fine features in the nuts. 
We investigated the water and chemical content in areca nut. 
% In the earlier growth and after maturation is much perceptible. 
These areas were more viscous in the early phases of ageing.
As the nuts matures, the fine characteristic white grains appear in specific areas.
Earlier, we observe in the Figure \ref{figure:2}\textbf{a} the saturate regions in the unripened nuts in contrast to ripened nuts.
In Figure \ref{figure:4}\textbf{e}, we show the evidence of fine features, which are responsible for nonlinear decomposition due to their different optical properties for both age groups of nuts.
These regions are enlarged for both the ageing groups.
The symbols represents the respective nuts of the Figure \ref{figure:3}\textbf{a} and Figure \ref{figure:3}\textbf{f}. 
%The white features are responsible for the edges and high-frequency signals are shown in Figure \ref{figure:2}\textbf{e}. 
In Figure \ref{figure:4}\textbf{f}, Figure \ref{figure:4}\textbf{g}, and Figure \ref{figure:4}\textbf{h}, we show the canny edge detection of unripened nuts and its zoomed in version of picking moisture content and some white grains.
Also, it is noticeable that the very small white micro grains are consolidated in the same image without any moisture unlike for unripened nuts shown in Figure \ref{figure:4}\textbf{i}, Figure \ref{figure:4}\textbf{j}, and Figure \ref{figure:4}\textbf{k} white consolidated particles are more. 
An attempt was made to identify the maturation growth and study the associated change in the alkaloid contents of two regions using MRI and mass spectrometer\cite{srimany2016developmental}. 
This study \cite{srimany2016developmental} helped us to hypothesise that at the final stage of maturity, the major quantity alkaloids are arecoline, arecaidine, and guvacoline get segregated in the brown region whereas guvacine gets to the white region of the areca nut.  

%Long-Range Anomalous Decay of the Correlation in Jammed Packings [PRL 2021 - Parisi et al.] - connection with the netowrk formation in areca nut.
%Community Detection in Quantum Complex Networks [PRX - Biamonte et al. 2014] - connection with DOS. 
% Between order and chaos James P. Crutchfield
%
The development of smartphone camera-based high-resolution measurement reveals dynamic heterogeneity of ageing for complex networks in areca nut from colour vision perspective.
It provides a unique coefficient from DOS to represent age and ageing together using a single number, average of them for less aged nuts and highly aged nuts are 4.9 and 3.8, respectively.
From the colour analysis, we found that molecules responsible for red are delocalised over the kernel of the nut.
Large number of peaks in the histogram supports high mAP values than the less peaks in the histogram.
The single peaks in the histogram of red channels for the less aged nuts corresponds to less mAP values.
An opposite trend is observed for the matured nuts, which indicates heterogeneity in delocalisation of red molecules due to new appearance of other coloured molecules.  
The average of red mAP, green mAP, and blue mAP are 1.02, 4.12, and 4.57 for less aged nuts and 2.12, 3.32, and 3.62 for matured nuts, respectively. 
It clearly indicates that variation in colour vision from one animal species to another will interpret different age for complex non-monochromatic networks of areca nuts.
We understood that the white fine features are responsible for the high-frequency signals in the Fourier analyses.
Furthermore, it opens new research in reproduction and growth such as, if transport of nutrients with maturity is a potential reason for peripheral localisation of DOS and associated with multitudes of physio-chemical changes in early development of progeny.
If a fruit looks aged from its external morphology as well as internal morphology, our method identifies the intrinsic similarities among the ageing networks without implementing any computationally expensive search algorithm. 
A diversity of ageing is strongly evident from the relative and absolute colour vision perspectives.
We have demonstrated a dynamic heterogeneity in the ageing-dependent density of states associated with the phenotypic expressions. 
%
%A typical trend of linearity between differently aged nuts are not followed like a typical periodic structure, which led us to calculate the density of each colour to identify their probability distribution using DOS. 
The complex model system of areca nut directs us to further investigate if the diverse responses are sensitively dependent on the initial physiological conditions and nonlinear chemical interactions by real time diffusion mechanism models.
Our method will be applicable to study a wide range of species and molecular studies and to develop diagnostics tools for health and food safety.

%Furthermore, a large number of areca nuts has been studied since 2017 to develop a low-cost tool to categorise areca nuts since the nut is considered to be addictive for humans and has increased adverse health effects, such as oral cancer. 
% 
% A typical trend of linearity between differently aged nuts are not followed like a typical periodic structure (See Supplementary Figure xx), which led us to calculate the density of each colour to identify their probability distribution using a density of states (DOS). 
% 
% In our case, it is also true, that complexity starts from the peripheral localisation, just like, the embryo development and cell proliferation in humans \cite{shahbazi2020mechanisms} is simple when the first step of cell division takes place with embryonic genome activation. 
% When it starts polarising and becomes a hollow cavity, the whole system will be more complicated, again, this complexity starts from the edges. 

%their ageing-related research, evolutionary psychology of complex pattern formation for phenotypic expression of age, the chemistry of health for areca nuts consumption, and computer vision for developing efficient algorithms.
\section*{METHODS}
\begin{figure*}[htp]
    \centering
    \includegraphics[width=0.5\textwidth]{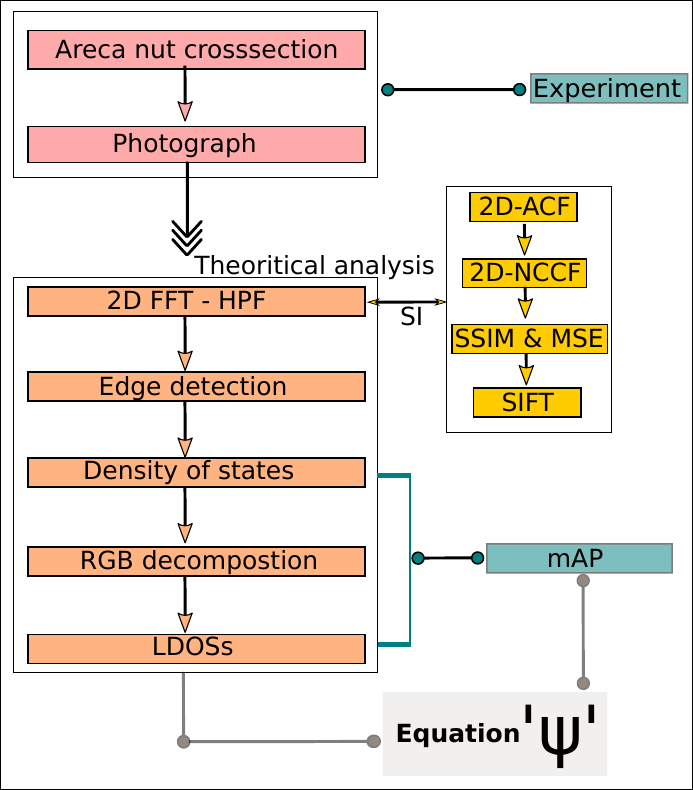}
    \caption{\textbf{Flowchart of the methods for paper.} Starting from experimental study to the total analysis. The $\Psi$ represents the total equation of this experiment, which is also unique alternate approaches to explain the solution for the problem.}
    \label{figure:5}
\end{figure*}

%We show the workflow of the study in Figure \ref{figure:5}.
This study followed the same protocol as in the Figure \ref{figure:5}.
OnePlus Technology (Shenzhen) Co Ltd's OnePlus 6T smartphone camera (16 megapixels, f/1.7, 25 mm (wide), 1/2.6", 1.22 m, PDAF, OIS) was used to create this image (Shenzhen, China).
All images were taken with same light, same height and same background. 
For Canny edge detection, 2D-ACF, structural similarity index, mean square error, and scale invariant feature transform, we have used grey scale images of original \cite{russ1990image}.
Since it is important because it is relatively easier to grasp the mathematics of it since it is a  single colour channel than multiple colour channels. 
Grey scale image also quantifies three dimensions image to two dimensions. rgb2grey converts RGB values to grey scale values by forming a weighted sum of the red, green, and blue components. The following equation is used to convert RGB to grey \cite{wu2007leaf} $0.2989 \times R + 0.5870 \times G + 0.1140 \times B$.
Structural similarity index matrix (SSIM) and mean square error (MSE) measurement has done to see the difference in differently aged groups. 
The detail discussion in Supplementary information table 1. 
We have also shown convolution of DOS and LDOSs of differently aged groups in one frame to understand the variations shown in supplementary Figure 6 and Figure 7.
We have extracted the special features of nuts using SIFT algorithm however it does not gave the best results as canny edge detection. 
SIFT started picking features outside the nut which is shown in supplementary Figure 8.
We have used Fiji ImageJ \cite{schindelin2012fiji} software to analyse canny edge detection. 
We found best detection at Gaussian smoothing $\sigma_f = 2$, low threshold is $2.5$ and high threshold is $7.5$. 
We have also tried different threshold value of Gaussian and kept low threshold is 2.5 and high threshold is 7.5 constant. 
We have also tried different Gaussian smoothing at $\sigma_f = 3$ and $\sigma_f = 1$ (See supplementary Figure 9). 
We have used Python programming to calculate plot-profile, 2D-FFT, RGB colour decomposition, and mAP. 
We have used Inkscape for data visualisation and vector drawing. 
Using inkscape, for Figure \ref{figure:2}\textbf{c} and Figure \ref{figure:2}\textbf{g} we added a 38.8\% transparent white box on the areca nut and layered it in between the edges and the nut.
The sample preparation method is discussed detailed in supplementary information Figure 10. 
We are deriving unique equation $\Psi$ for the explaining states of complexity in areca nut. 
\begin{equation}
    \Psi = (T)+(E)
\end{equation}
$\Psi$ is the complexity equation which is divided into two segments, theoretical study (T) and experimental study (E). 
The current analysis is solely theoretical, hence $E=0$ or $E$ is not taken into account in this study.
Initially, we used 2D FFT to extract high frequency signals from the original image sample $m \times n$ samples of areca nuts by blocking the frequency components below the cutoff frequency. 
Fourier transform will reflect the frequencies of periodic parts of the image.
Filtering is the process of removing undesirable frequencies using the inverse Fourier transform.
A filter is a matrix, and its components typically range from 0 to 1.
The frequency is permitted to pass if the component is 1, but it is cast away if the component is 0.
\begin{equation}
 \zeta_{HP}(u,v) = \Bigg\{\begin{cases}1, \ r(u,v)> F_{c} \\0, \ otherwise \end{cases}
\end{equation}
where, $r(u,v)= \sqrt{m^2+n^2}$ and $F_c$ is the cutoff frequency. 
Number of high frequency states are not clearly visible. 
Hence, we further analysed taking canny edge detection of original images and introduced the DOS.  
The density of states related to each image $I$ with countable pixels $P$ and their states $\xi$ can be defined as: 
\begin{equation} \label{eq:3}
 \xi(P) = \frac{1}{I} \sum_{i=1}^{N} \delta (P-P(ki)).
\end{equation}
The smallest allowed change in pixel $k$ of the edge detected pixel in an image of dimension $\mathcal{D}$ and length $l$ is $(\delta k)^\mathcal{D}$ $=$ $(\frac {2\pi}{l})^\mathcal{D}$. The density of states for continuous pixels are obtained within the $\lim_{l \to \infty}$, as
\begin{equation} \label{eq:4}
\xi(P)= \int_{\mathcal{D}} \frac{\mathcal{D}^k}{(2\pi)^\mathcal{D}} \delta (P-P(ki)). 
\end{equation}
Equivalently, the density of states can also be understood as the derivative of the microcanonical partition function $Z_m(P)$ (that is, the total number of states with pixels less than $P$ with respect to the pixels: 
\begin{equation} \label{eq:5}
\xi(P)=\frac {1}{I} \frac{dZ_m (P)}{dP} \delta (P-P(ki))
\end{equation}
%The number of states with pixels  $P^\prime$ (degree of degeneracy) is given by:
%\begin{equation} \label{eq:6}
 %   g(P^\prime)= \lim_{\delta P \to 0 } \int_{P^\prime}^{P^\prime+\delta P} \xi(P)dP
   % = \lim_{\delta P \to 0 } \xi(P^\prime) \delta P
%\end{equation}
The quantification from RGB decomposition, the edge detected states are the local density of states (LDOSs). 
This is given by the equation,  
\begin{equation} \label{eq:6}
 \chi(P,x)= \sum_{n} \bigg\{\mid \phi_n(x)^2 \mid \delta (P-P(ki))\bigg\}
 \end{equation}
The term $\mid\phi_n(x)^2\mid$ shows that each state contributes more in the regions where the density is high. 
An average over $x$ of this expression will replicate the usual formula for a DOS.
Since our model of study is non-homogeneous systems, the LDOSs play a major part to extract more information. 
For instance, $\chi(P,x)$ contains more information than $\xi(P)$ in equation \ref{eq:3} alone. 
In Figure \ref{figure:4}\textbf{a} and Figure \ref{figure:4}\textbf{b}, we show the total sum of LDOSs and compared with DOS. 
The equation of that is given by,
\begin{equation} \label{eq:7}
    \chi(P,x)= \sum_{i=1}^{n} \Bigg\{\sum_{n} \bigg(\mid \phi_n(x)^2 \mid \delta (P-P(ki))\bigg)\Bigg\}
\end{equation}
From the analysis we found that total DOS of a nut ($\xi(P)$) is not equal to sum of total LDOSs of its corresponding nuts $\chi(P,x)$.
Mathematically, equation \ref{eq:5} and equation \ref{eq:7} are not equal. 

The mean average precision ($\Upsilon$) is the arithmetic mean of the average precision values for an information retrieval system over a set of $n$ query topics. 
Where $\mu$ represents the average precision value for a given topic from the evaluation set of $n$ topics. 
From this we have also calculated standard deviation, which appraise the amount of variation or dispersion of a set of values. 
A low standard deviation $\sigma$ indicates that the values tend to be close to the mean of the set, while a high standard deviation indicates that the values are spread out over a wider range.
\begin{equation} \label{eq:8}
    \Upsilon_{DOS} = \frac{1}{n} \sum_{n} \mu_n
\end{equation}
Where, $\mu_n$ is DOS in the equation \ref{eq:3}. 
equation \ref{eq:8} becomes,
\begin{equation} \label{eq:9}
 \Upsilon_{DOS} = \frac{1}{n} \frac {1}{I} \frac{dZ_m (P)}{dP} \delta (P-P(ki))
\end{equation} 
Similarly, $\Upsilon_{LDOSs}$ is defined as, 
\begin{equation} \label{eq:10}
    \Upsilon_{LDOSs} = \frac{1}{n} \sum_{n} \mu_n
\end{equation} 
From equation \ref{eq:9}, equation \ref{eq:9} can be rewritten as 
\begin{equation} \label{eq:11}
    \Upsilon_{LDOSs} = \frac{1}{n} \sum_{n} \mid \phi_n(x)^2 \mid \delta (P-P(ki))
    \end{equation}
equation \ref{eq:9} and \ref{eq:11} are the complexity coefficients of $\Upsilon_{DOSs}$ and $\Upsilon_{LDOSs}$ respectively. 
The standard deviation is given by $\Omega$, equation becomes,
\begin{equation} \label{eq:12}
\Omega = \sqrt{ \frac{1}{N} \sum_{i=1}^n (x_i-\mu)^2}    
\end{equation}
Where $N$ is the total number of LDOSs, $x_i$ is the each LDOSs and $\mu$ is the mean average precision.
\\
\section*{Code availability}

The code used to generate the data from the experiments can be accessed from directly contacting KVC. 

\section*{Acknowledgements}

We would like to acknowledge the internal funding and resources of the Open Academic Research Council. 
We are grateful to the Council for giving a unique platform to perform publicly proposed research like this. 
Areca nuts used here are grown in a scientific farm of KVC, our sincere thanks to the public involved in maintaining the farm. 
We are also thankful to Rural Community Science Center (Sagara, Karnakata, India) for the samples. 
We are thankful to Dr Anjana Ramesh, Dr Pritam Pai, Jaishankar Hebbali, and Dr Abhilash Thendiyammal for their advice and initial supports for this work within the council.

\section*{Author contributions statement}

KVC proposed the problem and conducted the experiments. 
KVC and SG analysed the results and wrote the manuscript. 
MG provided critical advise on the manuscript structuring. 
SG designed and supervised the research. 
All authors reviewed the manuscript. 

\section*{Competing interests} 

Author MG is employed by the Thermo Fisher Scientific. 
Author KVC is employed by the International Centre for Nano devices. 
The remaining authors declare that the research was conducted in the absence of any commercial or financial relationships that could be construed as a potential conflict of interest.

\bibliography{mainref.bib}

\clearpage

\section*{SUPPLEMENTARY INFORMATION}

\section*{2D- Fast Fourier transform} Showing complexity patterns in  living organisms is an the essential phenomenon in our daily life. 
One way to identify the complex system in biology is the Fourier transform, states that the non periodic signals whose area under the curve is finite can also be represented into integrals of the sines and cosines after being multiplied by a certain weight. 
The Fourier transform has many wide applications that include image compression, filtering and image analysis.
Since we are dealing with digital images, we will be working with a discrete Fourier transform.
%We computed 2D-Fourier Transform using this formulae, taken from Matlab considering the Fourier term is sinusoidal. 
It includes three terms, spatial frequency, magnitude, and phase.
The spatial frequency directly relates with the position on the Fourier image. 
The magnitude of the sinusoidal directly relates with the brightness of that pixel in Fourier space. 
In our present study, we are concerned about digital images, so we consider a square image of size  $m \times n$ The 2D Fourier transform is given by
%The formula defines the discrete Fourier transform $Y$ of an $m$-by-$n$matrix $X$: 
\begin{equation}
Y_{p+1,q+1} =  \sum_{j=0}^{m-1} \sum_{k=0}^{n-1} \omega_m^jp \omega^kq_n X_{j+1,k+1}
\label{Sup equation:1}
\end{equation}
$\omega_m$ and $\omega_n$ are complex roots of unity.
Where, $\omega_m = e^\frac{-2\pi i}{m}$ and $\omega_n = e^\frac{-2\pi i}{n}$ is the imaginary unit. $p$ and $j$ are indices that run from $0$ to m-1, and $q$ and $k$ are indices that run from $0$ to n–1. This formula shifts the indices for $X$ and $Y$ by $1$ to reflect matrix indices. 

\begin{figure*}[htp]
    \centering
    \includegraphics[width=1\textwidth]{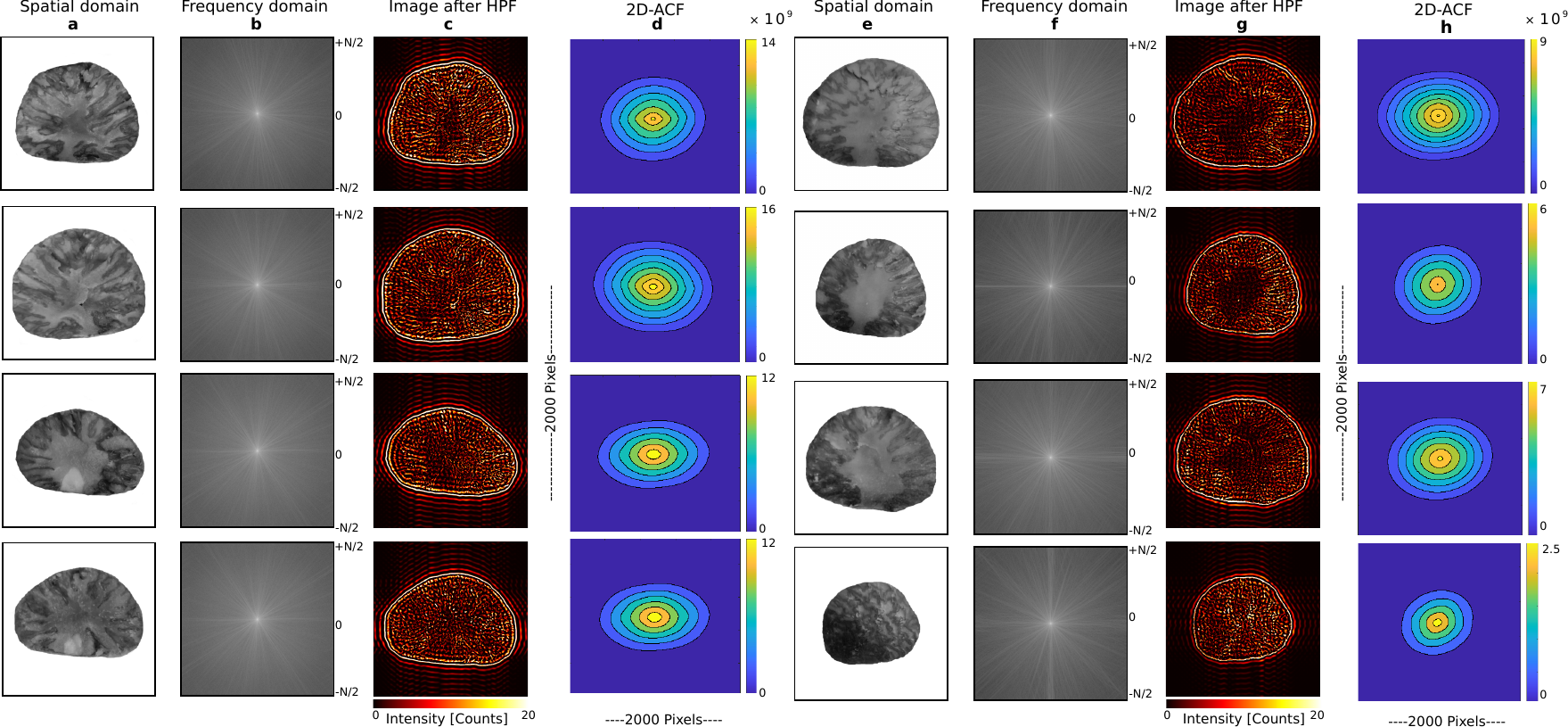}
    \caption{\textbf{2D Fast Fourier transform (2D-FFT) of ripened and unripened areca nuts}. \textbf{a}, Grey image of original unripened nuts in spatial domain. \textbf{b}, 2D FFT of corresponding grey image of nuts in \textbf{a}. \textbf{c}, image after high pass filter (HPF) from the frequency domain  \textbf{b}.  
    \textbf{d}, 2D-auto correlation of Grey image of original unripened nuts from \textbf{a}.
    \textbf{e}, Grey image of original ripened nuts in spatial domain. \textbf{f}, 2D FFT of corresponding grey image of nuts in \textbf{e}. \textbf{g}, image after high pass filter (HPF) from the frequency domain  \textbf{f}.  
    \textbf{h}, 2D-auto correlation of Grey image of original ripened nuts from \textbf{e}.}
    \label{Supplementary Fig:1}
\end{figure*}

In Figure \ref{Supplementary Fig:1}\textbf{a} and \ref{Supplementary Fig:1}\textbf{b}, the grey image of unripened and 2D-FFT images followed by grey image of ripened nut and its 2D-FFT in \ref{Supplementary Fig:1}\textbf{e} and \ref{Supplementary Fig:1}\textbf{f}. 
The centre bright in \ref{Supplementary Fig:1}\textbf{b} and \ref{Supplementary Fig:1}\textbf{f} shows the low frequency signals and the specific long stripes represents the edges. 
Areca nut has a strong edges outside and also morphological patterns has also a strong edges. 
It is clear that the frequency domain clearly explains there are complexity patters present in all the areca nuts. 
The basis functions are sine and cosine waves with increasing frequencies, i.e. $Y(0,0)$ represents the DC-component of the image which corresponds to the average brightness and $F(M-1,N-1)$ represents the highest frequency. 
The basis functions are sine and cosine waves with increasing frequency, That is $Y(0,0)$ represents the DC-component of the image which corresponds to the average brightness and $Y(N-1,M-1)$ represents the highest frequency.

\begin{figure*}[htp]
    \includegraphics[width=1\textwidth]{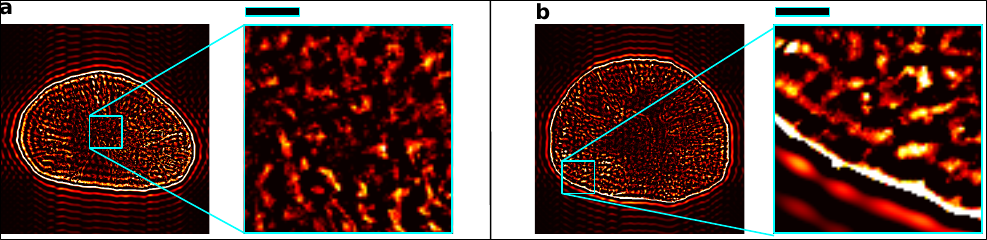}
    \caption{Detailed structure of high pass filter (HPF) in both ripened and unripened areca nuts. \textbf{a}, An exemplary unripened nut and its fine features. \textbf{b}, one particular ripened nut and its fine features. In \textbf{b}, it is also clearly shows the ringing artefacts.  
    All of them having same size of enlarged images with the fine features of (scale bar = 9.1 mm).}
    \label{Supplementary Fig:2}
\end{figure*}
%it is the first stage of maturation. With the naked eye we can say it has more white regions compared to brown. We took a FFT of it, more the signal is situated near f(0,0) , that is, it does not contain a high frequency signal. But if we can observe keenly, there are some linear lines passing out from f(0,0). Which means, there is a significant rate of change of the spatial frequencies in those directions in the real space image. In the later examples we are able to see how the transformation is happening with respect to maturation of the nut. The color bar represents the range from 0-1000 on both x and y axes. The color bar is normalised to the maximum magnitude of any spatial frequency in the image. The maximum frequency corresponds to the Nyquist frequency. Nyquist frequency for N samples is N/2. size returns the number of samples in each direction, so divide the size by 2 to get the frequency. This corresponds to the maximum frequency reliably expressible, which is the case where alternating pixels are off and on.The 0-1000 goes from -Nyquist to +Nyquist frequency.

\begin{figure*}[h!]
    \centering
    \includegraphics[width=0.9\textwidth]{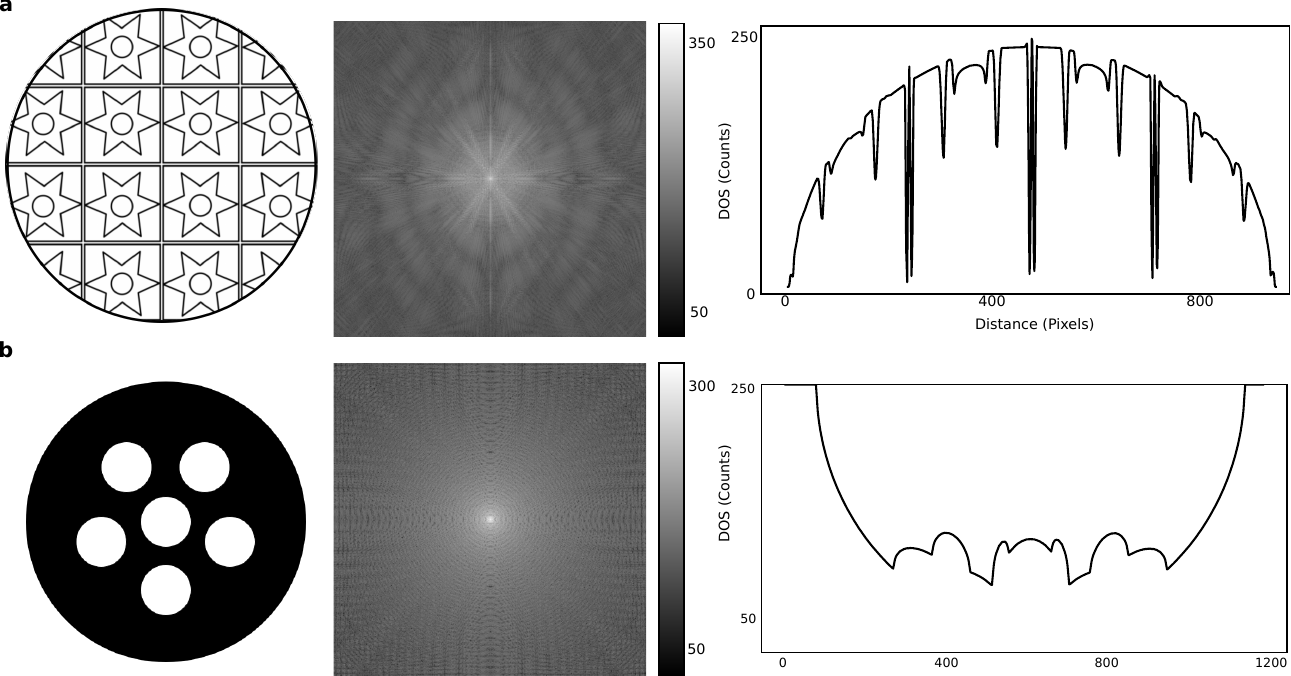}
    \caption{Reference crystalline pattern as control for FFT and DOS. \textbf{a}, Pattern with specific periodic patterns which is followed by its FFT and DOS. FFT and DOS clearly shows the periodic signals.
    \textbf{b}, exemplary crystal structure lattice patterns without any defects, followed by FFT and DOS. The periodic signals are visible here too without any complex patterns. }
    \label{Supplementary Fig:3}
    \end{figure*}

\section*{2D-Auto-correlation} 
In Figure \ref{Supplementary Fig:1}\textbf{d} and \ref{Supplementary Fig:1}\textbf{h}, the auto-correlation of unripened and ripened nuts shown respectively. 
The auto-correlation describes how good an image correlates with in all possible pixels. 
A correlation in object space "reduces" to a multiplication infrequency space.
The 2-D auto-correlation of an $m$ by $N$ matrix, $X$, and $P$ by $Q$ matrix, $H$, is a matrix, $C$, of size $M + P - 1$ by $N + Q - 1$ Its elements are given by
\begin{equation}
C(k, l)= \sum_{m = 0}^{m - 1} \sum_{n=0}^{n - 1} X(m,n) H(m - k,n - l), -(P - 1)\leq k \leq {M - 1}, -{(Q - 1)} \leq l \leq {N - 1}.
\label{Sup equation:2}
\end{equation}
Here, $ -(P-1) \leq k \leq M-1, -(Q-1) \leq l \leq N-1$. 
Noting that left hand side $C(k,l)$ yields 2D-ACF if $k=l$ otherwise it gives 2D-cross correlation and the bar over H denotes complex conjugation.
The image's centre has the highest correlation.
That is, the white region produces the highest power spectrum at the centre, as illustrated by the yellow colour.
Because there is a background in it, the background pixel is zero.
The blue intensity is shown to be zero in the 2D ACF contour graphic.
Similarly for the second set of sample in Figure \ref{Supplementary Fig:1}\textbf{h} (differently aged areca nut samples) auto correlation is not maximum. 
From both sets of samples, ACF images contains elliptical patterns which are distributed non-homogeneously over the entire area of 1200 $\times$ 1200 pixels. 
The ACF is expected to reveal the shape anisotropy (orientation and shape ratio) of ellipses through the contour lines which surround the origin closely.
The contour plot displayed in figure confirms and proves there is no proper auto-correlation in the eight areca nut samples. 
\\
\section*{2D-Normalised cross-correlation} 
\begin{figure*}[htp]
    \centering
    \includegraphics[width=1\textwidth]{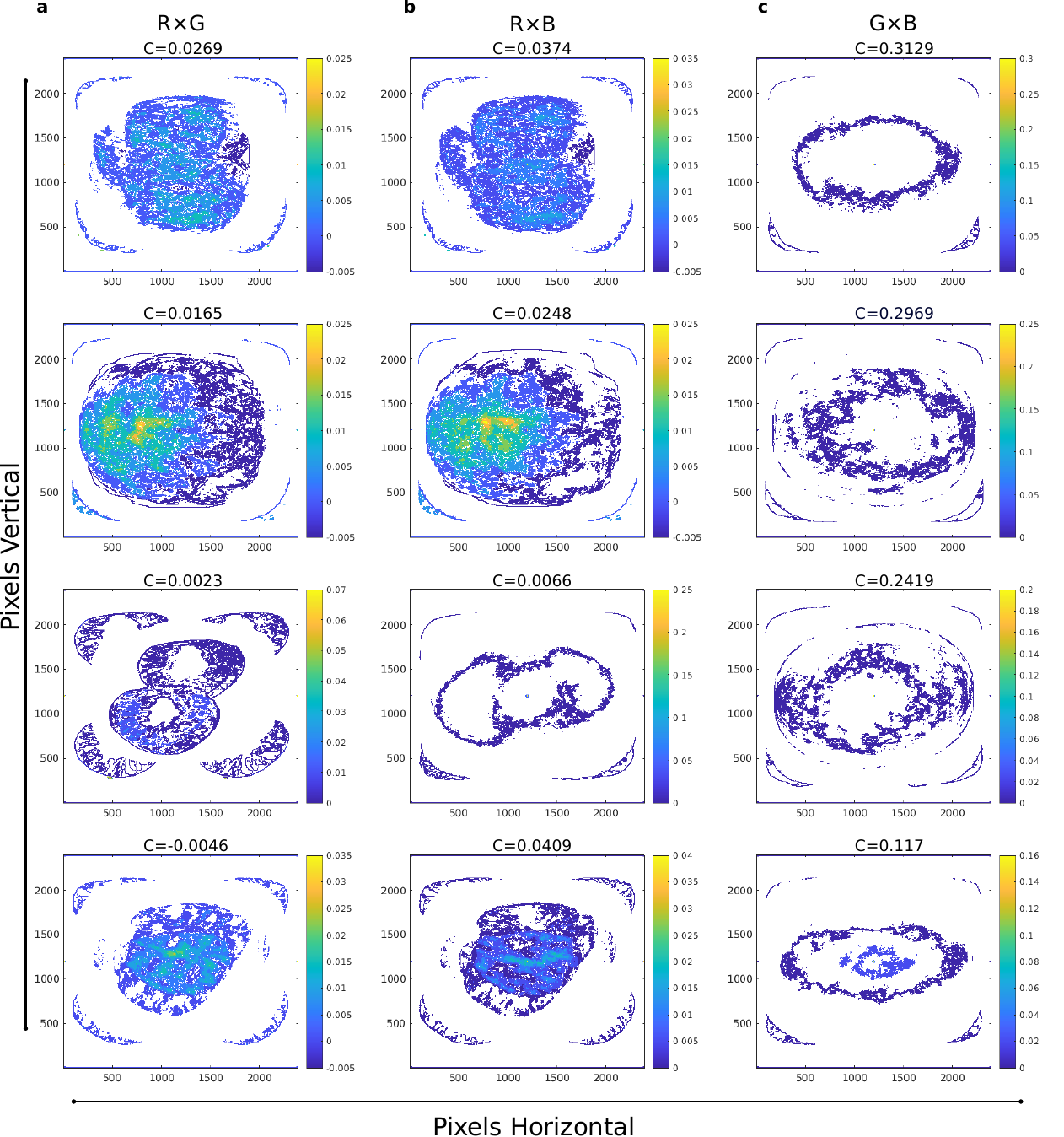}
    \caption{\textbf{2D-Normalised cross-correlation (2D-NCC) of four unripened nuts LDOSs.} All four nuts are analysed one to one combinations of the corresponding LDOSs channels. \textbf{a}, R$\times$G is analysed 2D-NCC of unripened nuts of LDOSs. 
    \textbf{b},  R$\times$B is analysed 2D-NCC of unripened nuts of LDOSs. \textbf{c}, G$\times$B are the one to one analysed 2D-NCC of unripened nuts LDOSs.
    The 2D-NCC coefficient in all the combinations \textbf{\textit{C}}, obtained from Equation \ref{Sup equation:3}. The value is of 2D-NCC coefficient is 1 iff R$\times$R, G$\times$G, and B$\times$B. 
    The colour bar [counts] in each 2D normalised cross correlation image matrix represents the maximum and minimum correlations. 
    All combinations are not close to 1. 
    Hence all combinations of LDOSs are not correlated. 
    Here white regions represents zero signal (matrix elements)}
    \label{Supplementary Fig:4}
\end{figure*}

\begin{figure*}[htp]
    \centering
    \includegraphics[width=1\textwidth]{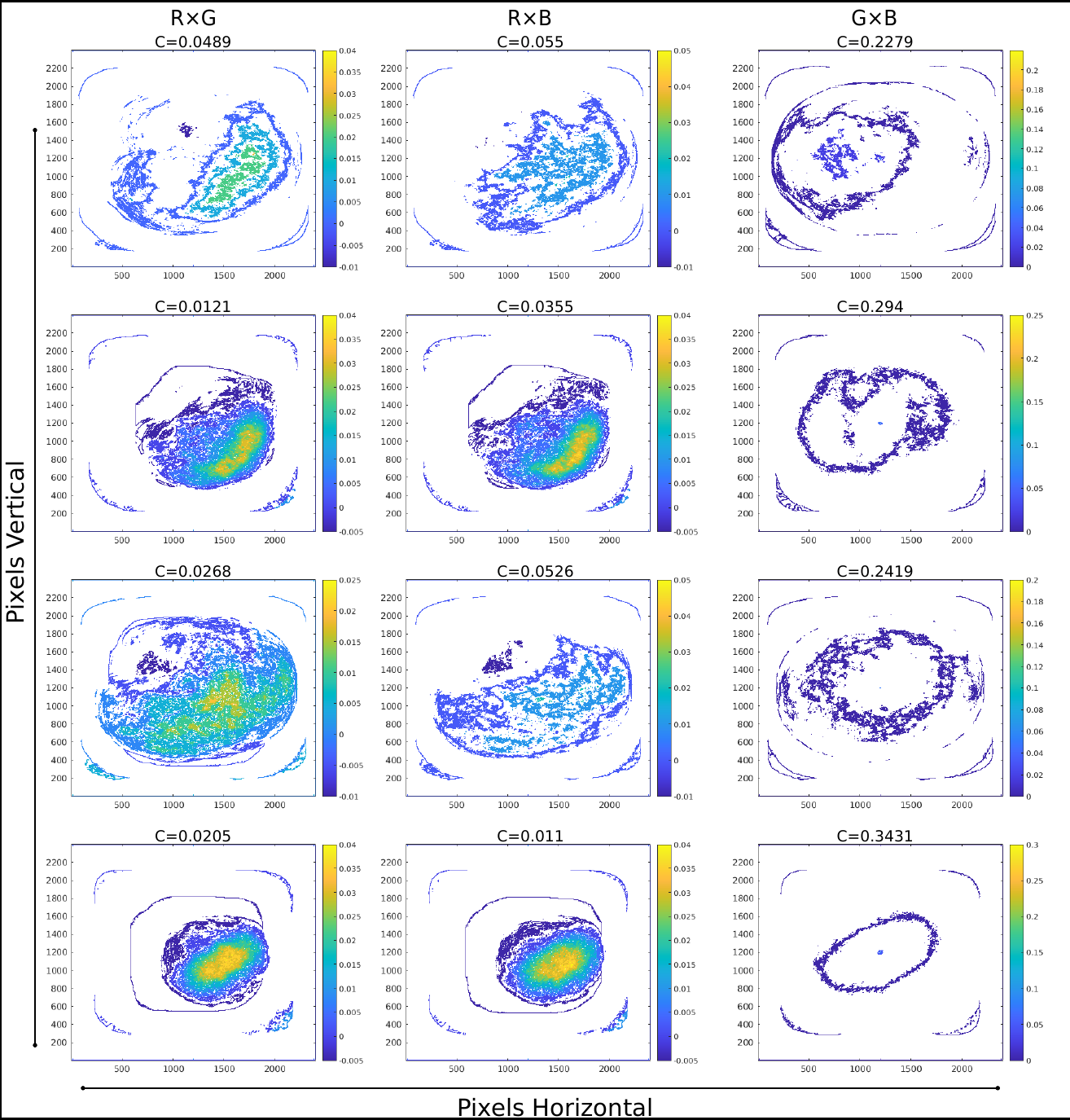}
    \caption{\textbf{2D-Normalised cross-correlation (2D-NCC) of  four ripened nuts LDOSs.} 
    All four nuts are analysed one to one combinations of the corresponding LDOSs channels. 
    \textbf{d}, R$\times$G is analysed 2D-NCC of unripened nuts of LDOSs. 
    \textbf{e},  R$\times$B is analysed 2D-NCC of unripened nuts of LDOSs. 
    \textbf{f}, G$\times$B are the one to one analysed 2D-NCC of unripened nuts LDOSs.
    The 2D-NCC coefficient in all the combinations \textbf{\textit{C}}, obtained from Equation \ref{Sup equation:3}. The value is of 2D-NCC coefficient is 1 iff R$\times$R, G$\times$G, and B$\times$B. The colour bar [counts] in each 2D normalised cross correlation image matrix represents the maximum and minimum correlations. All combinations are not close to 1. Hence all combinations of LDOSs are not correlated. Here white regions represents zero signal (matrix elements)}
    \label{Supplementary Fig:5}
\end{figure*}

We did the LDOSs 2D-normalised cross-correlation (2D-NCC) to see is there any strong correlation between the LDOSs. \ref{Supplementary Fig:4} and \ref{Supplementary Fig:5} tells us the 2D-NCC  with the correlation coefficient value $C$.
\begin{equation}
\xi_{u,v} =  \frac{\sum_{xy} [f(x,y)-f_{uv}] [t(x-u,y-v)-t]}{[\sum_{xy}[f(x,y)-f_{uv}]^2 \sum_{xy} [t(x-u,y-v)-t]^2]^{0.5}}
\label{Sup equation:3}
\end{equation}
where, $f$ is the image, $t$ is the mean of the template, $f_{u,v}$ is the mean of $f(x,y)$ in the region under the template.
This equation, calculate cross-correlation in the spatial or the frequency domain, depending on size of images then calculate local sums by precomputing running sums and uses local sums to normalise the cross-correlation to get correlation coefficients.

The correlation coefficient $\textbf{C}$, tells us about the strength and direction of the linear relationship between $x$ and $y$. 
We did LDOSs cross correlation to see if there is any linear relationship between two image metrics.
Three types of image metrics are compared R $\times$ G, R $\times$ B and G $\times$ B for both LDOSs of unripened and ripened nuts.
But the results shows there is a no significant linear relationship between $x$ and $y$. 
It concludes that the correlation coefficient is not significantly different from zero (it is close to zero), we say that correlation coefficient is “not significant”. 
\\
\section*{Structural similarity index matrix and mean square error}
We used two types of analysis to understand the similarities between the images: structural similarity index matrix (SSIM) and mean square error (MSE)\cite{lee1980digital,1284395}.
For the two images we are comparing, MSE will calculate the mean square error between each pixel.
SSIM, on the other hand, will work in the opposite direction.
SSIM will look for similarities within pixels, it will look for similar pixel density values.  
It is known that \cite{lee1980digital} the higher the MSE the less similar they are. 
SSIM on the other side puts everything in a scale between -1 to +1. 
A value of +1 means they are very similar and a value of -1 means they are very different. 
It is because the SSIM values calculates luminance, contrast and structure. 
The three components are combined to yield an overall similarity measure. 
\setlength{\arrayrulewidth}{0.5mm}
\setlength{\tabcolsep}{10pt}
\renewcommand{\arraystretch}{1.5}

\begin{table}[h]
\centering
\begin{tabular}{|p{2cm}|p{2cm}|p{2cm}|p{2cm}|p{2cm}|}
%\multicolumn{5}{|c|}{SSIM and MSE analysis} \\
\hline
Comparison & SSIM of Unripened nuts & SSIM of Ripened nuts & MSE of Unripened nuts $\times$ $10^3$ & MSE of ripened nuts $\times$ $10^3$ \\ 
\hline
1 $\times$ 1 & 1 & 1 & 0 & 0 \\ 
\hline
1 $\times$ 2 & 0.7074 & 0.7349 & 1.4595 & 5.1681 \\
\hline
1 $\times$ 3 & 0.7405 & 0.8028 & 0.7227 & 1.6394\\
\hline
1 $\times$ 4 & 0.7422 & 0.6447 & 3.066e & 9.0181\\
\hline
2 $\times$ 3 & 0.7157 & 0.7695 & 1.0734 & 3.5668\\
\hline
2 $\times$ 4 & 0.7178 & 0.7825 & 1.3402 & 3.7476\\ 
\hline
3 $\times$ 4 & 0.7807 & 0.6847 & 2.3117 & 7.2456\\ 
\hline
\end{tabular}
\caption{Comparison of SSIM and MSE analysis. 
The first row represents the number of nuts compared with each other. 1$\times$1 represents the first areca nut compared with first and so on.}
    \label{table:1}
\end{table}

The structural similarity index and mean square error are calculated with original unripened and ripened nuts. 
We compared the different combinations of two areca nuts in the order showed in Table \ref{table:1}. 
The image number 1 to 4 is in the same order showed in Figure \ref{Supplementary Fig:1}\textbf{a} and Figure \ref{Supplementary Fig:1}\textbf{e}. 
The value of SSIM gives 1 when both images are identical and it the similarity's deceased when it started comparing to non identical areca nut sample. 
To understand the MSE, it is important to note that a value of 0 for MSE indicates perfect similarity. Here, our comparing 1 $\times$ 1 for both resulting in 0.  
A value greater than one implies less similarity and will continue to grow as the average difference between pixel intensities increases as well. 
The highest error in the images is 1$\times$4 combination for both ripened and unripened nuts. 
It is clear that all of the photos are non linear, implying that there is a structural difference between the eight areca nuts in terms of luminance, contrast, and structure comparison.

\begin{figure*}[htp]
   \centering
   \includegraphics[width=0.7\textwidth]{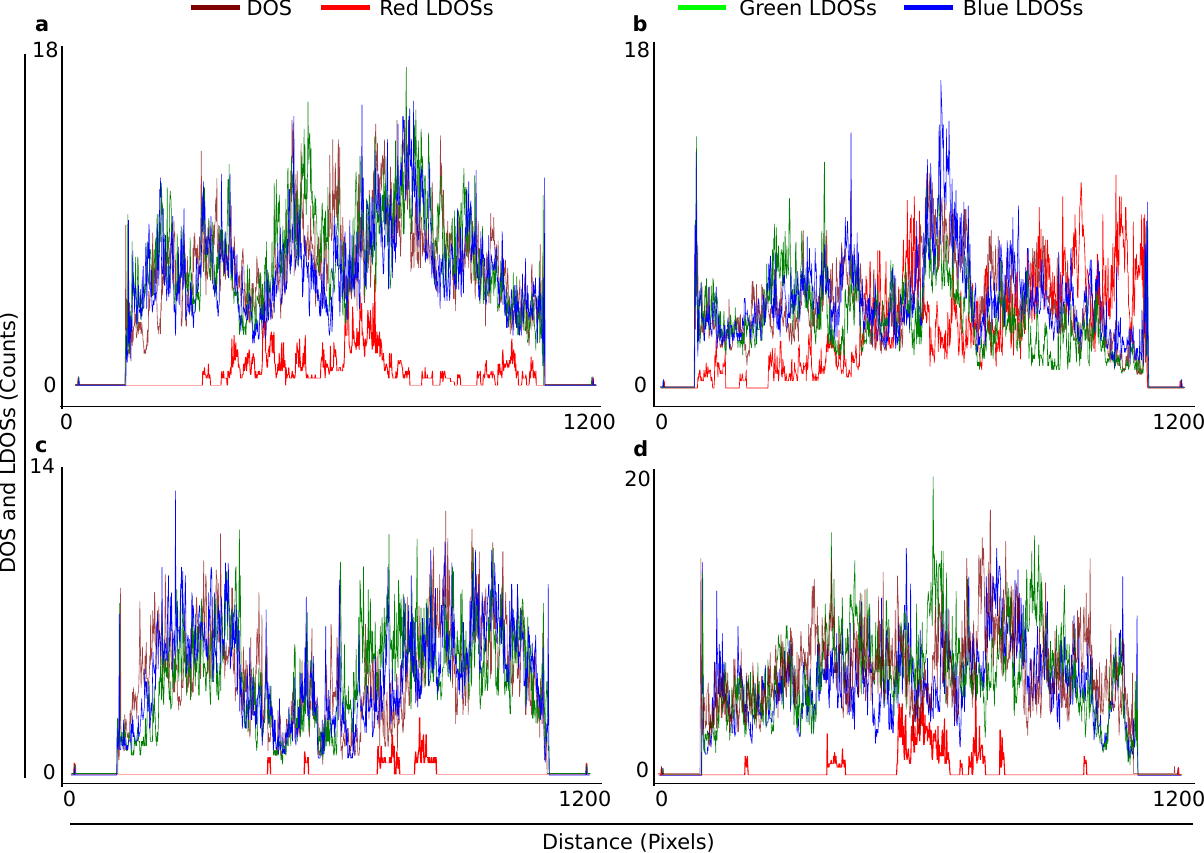}
    \caption{DOS and LDOSs convolution plot of unripened nuts which has a common x and y axis. 
    \textbf{a}, convolution of DOS with LDOSs plot of first unripened nut.
   \textbf{b}, convolution of DOS with LDOSs plot of second unripened nut. 
   \textbf{c}, convolution of DOS with LDOSs plot of third unripened nut and \textbf{d}, convolution of DOS with LDOSs plot of fourth unripened nut}
   \label{Supplementary Fig:6}
\end{figure*}

\begin{figure*}[htp]
   \centering
   \includegraphics[width=0.7\textwidth]{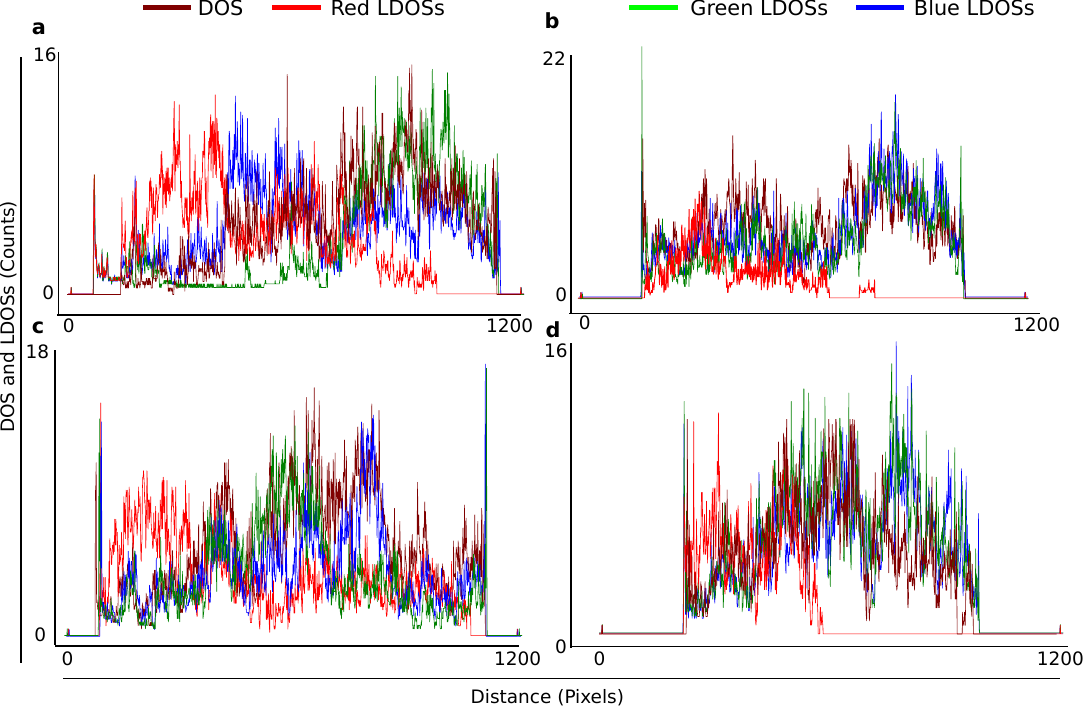}
   \caption{DOS and LDOSs convolution plot of ripened nuts which has a common x and y axis. \textbf{a}, convolution of DOS with LDOSs plot of first ripened nut.
   \textbf{b}, convolution of DOS with LDOSs plot of second ripened nut. \textbf{c}, convolution of DOS with LDOSs plot of third ripened nut and \textbf{d}, convolution of DOS with LDOSs plot of fourth ripened nut}
   \label{Supplementary Fig:7}
\end{figure*}

\section*{Scale invariant feature transform (SIFT)} 
% Initially, we were curious to see how areca nuts compared to SIFT.
We used SIFT to extract the most important characteristics from each of the eight areca nuts.
\begin{figure*}[h]
    \centering
    \includegraphics[width=1\textwidth]{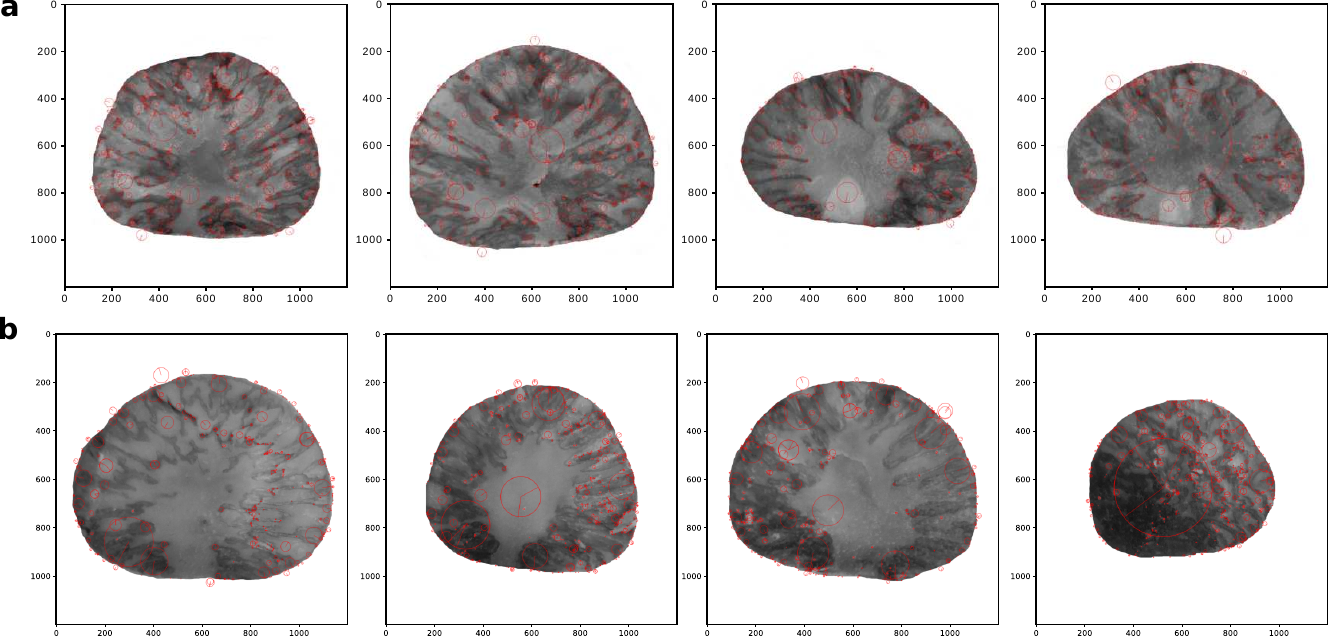}
    \caption{Scale invariant feature transform (SIFT) of unripened and ripened areca nuts. \textbf{a}, SIFT's of unripened areca nuts showing major Key points in the original Grey scale images.
    \textbf{b}, SIFT's of ripened areca nuts showing major Key points in the original grey scale images. 
    Interest points detected from all eight areca nut images. 
    The red arrows at the centres of the circles illustrate the orientation estimates obtained from peaks in local orientation histograms around the interest points.}
    \label{Supplementary Fig:8}
\end{figure*}

The SIFT key points matching is displayed in Figure \ref{Supplementary Fig:8} does not satisfy us. 
It is primarily picking key points from the edges.
Excellent properties of white drain chemicals of areca nut were expected from SIFT.
The LDOSs model produced superior outcomes.

\section*{Canny edge detection} 
The canny edge detection was observed at different values of $\sigma$, low and high thresholds. 
Edges are not well detected in all combinations, and there are more errors in edge selection.  
\begin{figure*}[htp]
    \centering
    \includegraphics[width=1\textwidth]{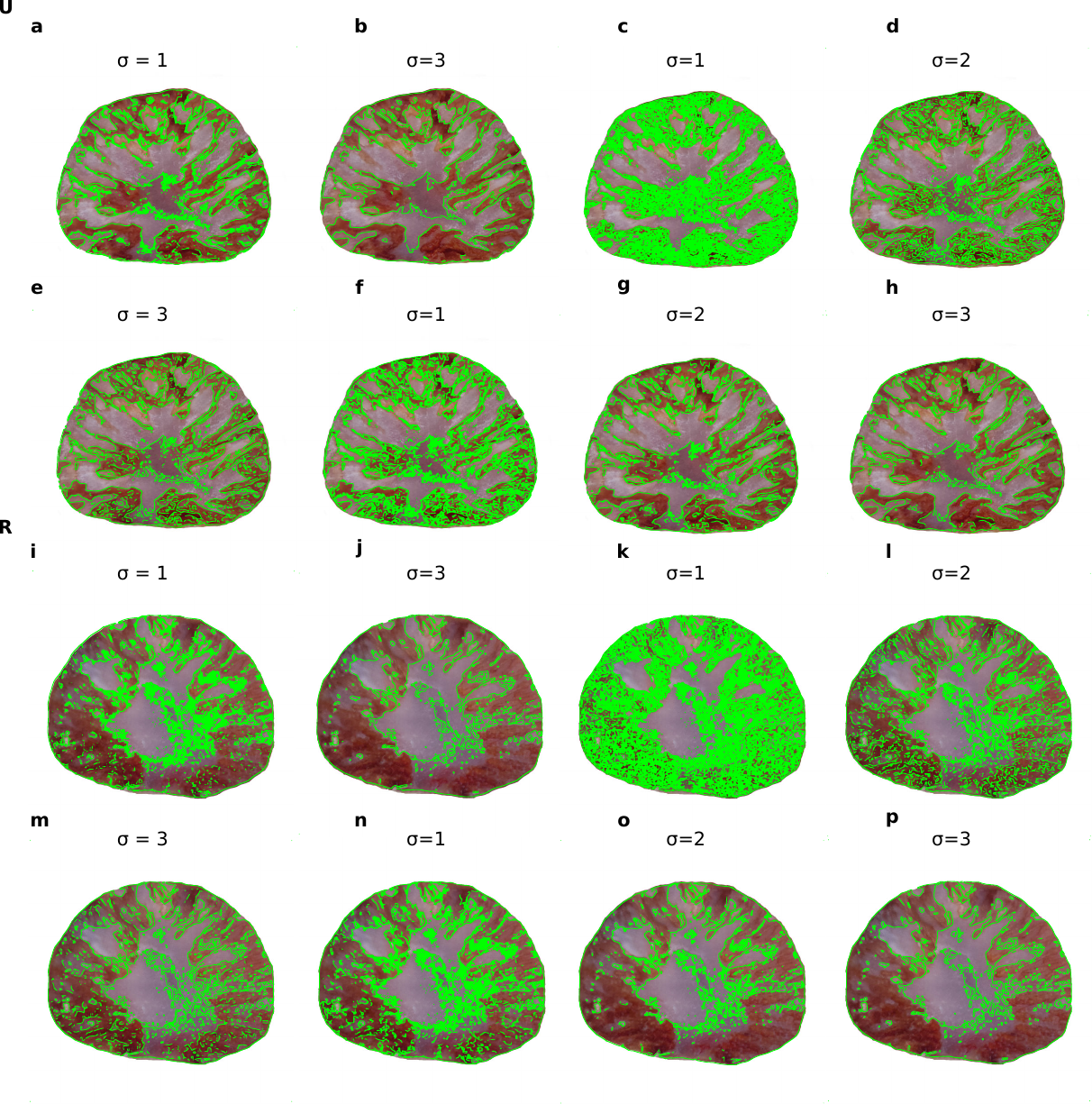}
    \caption{\textbf{Extended version of Canny edge detection with variations in Gaussian radius '$\sigma$', high, and low threshold's (HT and LT). 
    \textbf{U}, signifies one specific areca nuts and its analysis with different thresholds of unripened nuts and correspondingly, \textbf{R}, signifies the ripened nuts.} 
    \textbf{a}, evidence of Gaussian radius $\sigma$ is 1, LT is $2.5$, and HT is $3.5$. 
    \textbf{b},  evidence of Gaussian radius $\sigma$ is 3, LT is $2.5$, and HT is $3.5$. 
    \textbf{c}, evidence of Gaussian radius $\sigma$ is 1, LT is $2.5$, and HT is $1$. 
    \textbf{d}, evidence of Gaussian radius $\sigma$ is 2, LT is $2.5$, and HT is $1$. 
    \textbf{e}, evidence of Gaussian radius $\sigma$ is 3, LT is $2.5$, and HT is $1$. 
    \textbf{f}, evidence of Gaussian radius $\sigma$ is 1, LT is $1$, and HT is $3.5$. 
    \textbf{g}, evidence of Gaussian radius $\sigma$ is 2, LT is $1$, and HT is $3.5$. 
    \textbf{h}, evidence of Gaussian radius $\sigma$ is 3, LT is $1$, and HT is $3.5$. 
    Similarly, for ripened nuts,\textbf{i}, evidence of Gaussian radius $\sigma$ is 1, LT is $2.5$, and HT is $3.5$. \textbf{j},  evidence of Gaussian radius $\sigma$ is 3, LT is $2.5$, and HT is $3.5$. 
    \textbf{k}, evidence of Gaussian radius $\sigma$ is 1, LT is $2.5$, and HT is $1$. 
    \textbf{l}, evidence of Gaussian radius $\sigma$ is 2, LT is $2.5$, and HT is $1$. 
    \textbf{m}, evidence of Gaussian radius $\sigma$ is 3, LT is $2.5$, and HT is $1$. 
    \textbf{n}, evidence of Gaussian radius $\sigma$ is 1, LT is $1$, and HT is $3.5$. 
    \textbf{o}, evidence of Gaussian radius $\sigma$ is 2, LT is $1$, and HT is $3.5$. 
    \textbf{p}, evidence of Gaussian radius $\sigma$ is 3, LT is $1$, and HT is $3.5$.}
    \label{Supplementary Fig:9}
    \end{figure*}
In Figure \ref{Supplementary Fig:9}, all combinations of $\sigma$ shown and their accuracy in piking up the edges are very poor.
The best example is in Figure \ref{Supplementary Fig:5}\textbf{c} and \ref{Supplementary Fig:5}\textbf{e}, when the $\sigma$ is low the canny algorithm starts picking fake edges. 
When $\sigma$ is high it will not pick may edges.
The most effective combination is found in the main text.
The sample preparation procedure entails selecting several age groups of areca nuts from the same tree with the same progeny. 
The smoothly cross section was made from the knife shown in Figure \ref{Supplementary Fig:10}. 
The thickness of the sharp edge side was measured from screw gauge. 
The thickness of the sharp edge is  1.2 mm. 
\begin{figure*}[h]
    \centering
    \includegraphics[width=1\textwidth]{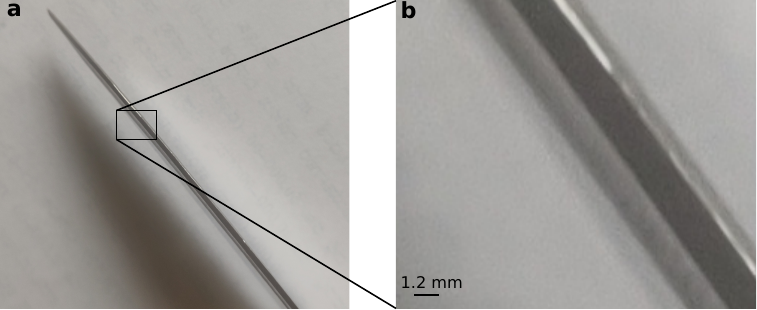}
    \caption{Knife used to cross section the areca nut.
    \textbf{a}, Top view of the knife. 
    \textbf{b}, Zoom in version of top view of the knife where the areca nut cross section was made. The thickness of the sharp edge is 1.2 mm.}
    \label{Supplementary Fig:10}
\end{figure*}

%\begin{figure*}[h]
 %   \includegraphics[width=0.6\textwidth]{archive/FigureSI/FigureS11.pdf}
%    \caption{\textbf{Gray-Scott Model of a Reaction-Diffusion System}.\textbf{a}, An exemplary diffusion model at the time 10s and its DOS. \textbf{b}, diffusion model at time 60s and its corresponding DOS.}
 %   \label{Supplementary Fig:11}
%\end{figure*}

\subsection*{Newton's colour triangle}
We have used Newton's colour triangle to identified the wavelengths of the primary colours (RGB) \cite{smits1999rgb, harkness2006colour}. 
We have used open access website which was utilised in the paper\cite{kuntzleman2016teaching} (https://academo.org/demos/wavelength-to-colour-relationship/) to convert RGB value to its respective wavelength. 
In Figure \ref{figure:3}\textbf{e} and Figure \ref{figure:3}\textbf{j}, the red, green, and the blue values are 255.  

% \clearpage
% \bibliography{SI.bib}

\end{document}